\documentclass[12pt,a4paper]{article}
\usepackage{epsfig}

\newcommand{\captionfonts}{\small}

\makeatletter  
\long\def\@makecaption#1#2{%
  \vskip\abovecaptionskip
  \sbox\@tempboxa{{\captionfonts #1: #2}}%
  \ifdim \wd\@tempboxa >\hsize
    {\captionfonts #1: #2\par}
  \else
    \hbox to\hsize{\hfil\box\@tempboxa\hfil}%
  \fi
  \vskip\belowcaptionskip}
\makeatother   

\begin{document}
\title{\textbf{Modelling of dislocation generation and interaction
    during high-speed deformation of metals.}\thanks{Presented at the
    Third Workshop on High-speed Plastic Deformation, Hiroshima
    Institute of Technology, August 2000.}}

\author{\textsc{J. Schi{\o}tz$^{1}$, T. Leffers$^2$ and B. N. Singh$^2$}\\
  $^1$ Center for Atomic-scale Materials Physics (CAMP)\\ and
  Department of Physics, Technical University of Denmark,\\ DK-2800 Lyngby.\\
  $^2$ Materials Research Department, Ris{\o} National Laboratory,\\
  DK-4000 Roskilde.}

\date{August 2, 2000}
\maketitle

\begin{abstract}
  Recent experiments by Kiritani \emph{et al.} \cite{KiSaKiArOgArSh99}
  have revealed a surprisingly high rate of vacancy production during
  high-speed deformation of thin foils of fcc metals.  Virtually no
  dislocations are seen after the deformation.  This is interpreted as
  evidence for a dislocation-free deformation mechanism at very high
  strain rates.
  
  We have used molecular-dynamics simulations to investigate
  high-speed deformation of copper crystals.  Even though no
  pre-existing dislocation sources are present in the initial system,
  dislocations are quickly nucleated and a very high dislocation
  density is reached during the deformation.
  
  Due to the high density of dislocations, many inelastic interactions
  occur between dislocations, resulting in the generation of
  vacancies.  After the deformation, a very high density of vacancies
  is observed, in agreement with the experimental observations.  The
  processes responsible for the generation of vacancies are
  investigated.  The main process is found to be incomplete
  annihilation of segments of edge dislocations on adjacent slip
  planes.  The dislocations are also seen to be participating in
  complicated dislocation reactions, where sessile dislocation
  segments are constantly formed and destroyed.\\
\end{abstract}

\section{Introduction}
\label{sec:intro}

In a recent paper, Kiritani \emph{et al.} reported that a large number
of vacancies were produced during high-speed heavy plastic deformation
of thin foils of fcc metals \cite{KiSaKiArOgArSh99}.  They observed a
large density of stacking-fault tetrahedra but very low dislocation
densities in the foils after deformation.  As a possible explanation,
they propose a dislocation-free deformation mechanism.

In this paper we investigate the deformation processes in metals under 
very high strain rates.  Our approach is that of Molecular Dynamics
(MD), where Newton's second law is solved numerically for all the
atoms in the system under consideration.  The main advantage of this
method is that no \emph{a priori} assumptions are made about the
deformation mechanism.  Structures such as dislocations appear
automatically from the motion of the atoms.  The simulation thus
becomes ``unbiased'', i.e.\ the result is independent of our prior
assumptions about the active deformation mechanisms.

We have chosen to simulate the deformation processes in systems that
are initially dislocation-free (i.e. without conventional dislocation
sources such as Frank-Read sources).  We start with single crystals in
order to avoid dislocation emission from grain boundaries.
Simulations of high-speed deformation of polycrystalline
(nanocrystalline) material have been reported elsewhere
\cite{ScDiJa98,ScVeDiJa99,SwCa98}.  Although many metals were studied
experimentally by Kiritani et al.\cite{KiSaKiArOgArSh99}, we have
chosen to focus our simulations on one metal: copper.

\section{Simulation methods and setup}
\label{sec:simul}

In order to get meaningful results from a molecular dynamics
simulation, it is essential that the interatomic forces used in the
simulation are a good approximation to the forces the atoms would
experience in a real physical system with the same atomic
configuration.  We use the effective medium theory (EMT)
\cite{JaNoPu87,JaStNo96} to describe the interatomic interactions, as
the EMT potential for copper has been well tested.  

The simulated systems were single crystals.  Two different system
sizes were used, one with approximately 95\,000 atoms and one with
approximately 765\,000 atoms.  The linear dimensions of the two
systems are approximately 10.5 nm and 21 nm.  In both cases periodic
boundary conditions were applied to the simulation, so the system
effectively becomes an infinite crystal.

In the smaller system four vacancies were created in the initial
configuration, by removing four random atoms.  They were introduced in
the unlikely case that vacancies were somehow actively participating
in the deformation.  The system is deformed along the [521] direction
at a temperature of 450 K.  We have chosen this temperature since we
consider it likely that significant local heating is occurring in the
experiments due to the high strain rate.

As the preexisting vacancies in the smaller system did not participate 
in the deformation, no vacancies were introduced in the larger
system.  It was deformed along the [$\bar{1}05$] direction at the same
temperature.

The systems were deformed at a constant strain rate, while keeping the
stress in the transverse direction approximately constant. This
procedure is further described in ref.\ \cite{ScVeDiJa99}.  In all
cases a strain rate of $10^9 s^{-1}$ was used (compared to
approximately $10^5 s^{-1}$ in the experiments).  To keep the
temperature of the system approximately constant during the
simulation, Langevin dynamics \cite{AlTi87} are used, i.e.\ a friction
and a fluctuating force are added to the equations of motion of the
atoms.  We use a timestep of 5\,fs, safely below the value where the
dynamics becomes unstable.

During the simulations, configurations were rapidly quenched, and the
local crystal structure was identified by common neighbor analysis
\cite{JoAn88,ClJo93}.  This was used to generate plots where all atoms
except atoms at the dislocation cores were made invisible, allowing
the visualization of the dislocation structures.

\section{Results}
\label{sec:results}

\begin{figure}[tp]
  \begin{center}
    \newcommand{\subfig}[1]{\raisebox{0.43\linewidth}[0pt][0pt]{\makebox[0pt][l]{\hspace*{1em}#1}}}%
    \subfig{(a)}%
    \epsfig{file=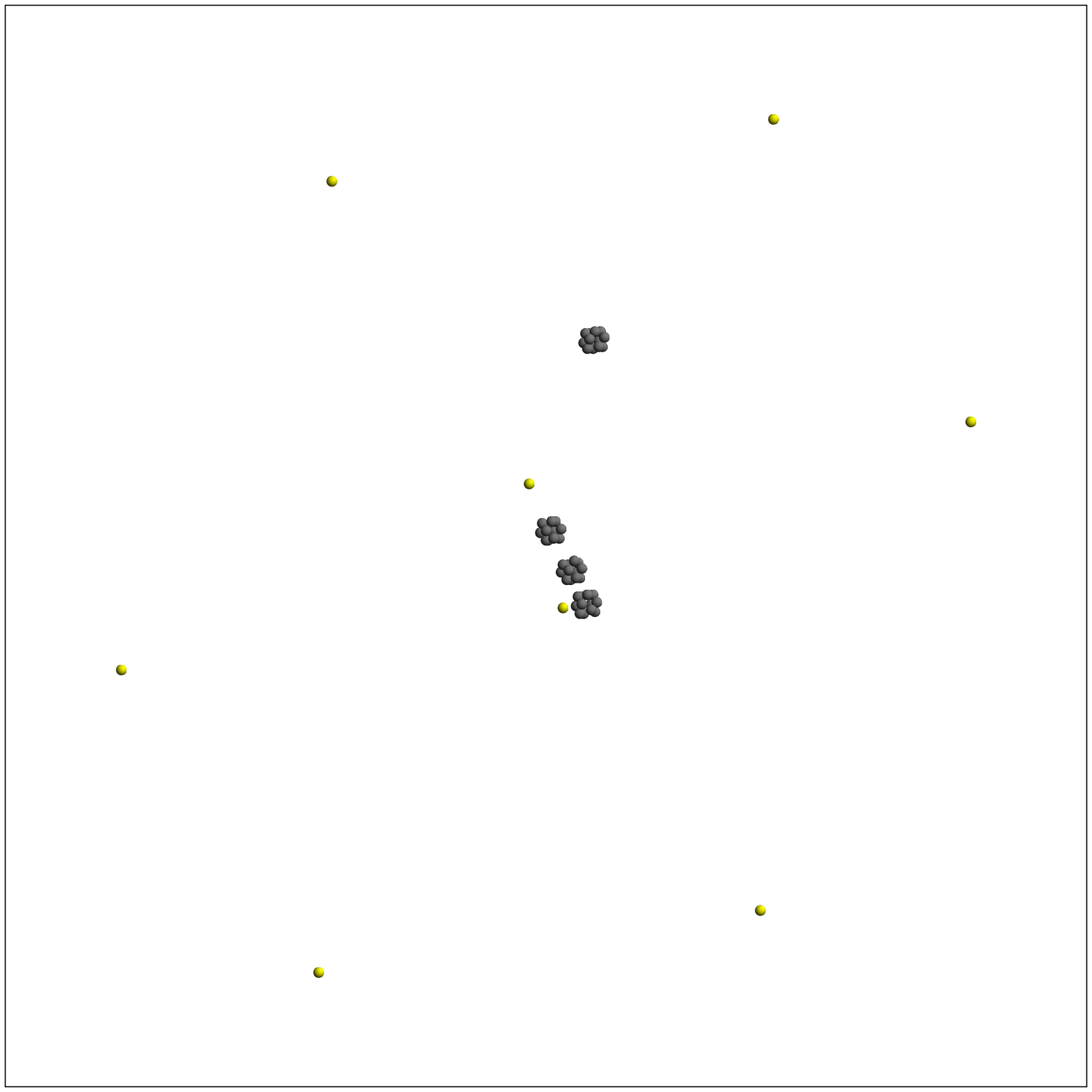, width=0.48\linewidth, clip=}\hfill
    \subfig{(b)}%
    \epsfig{file=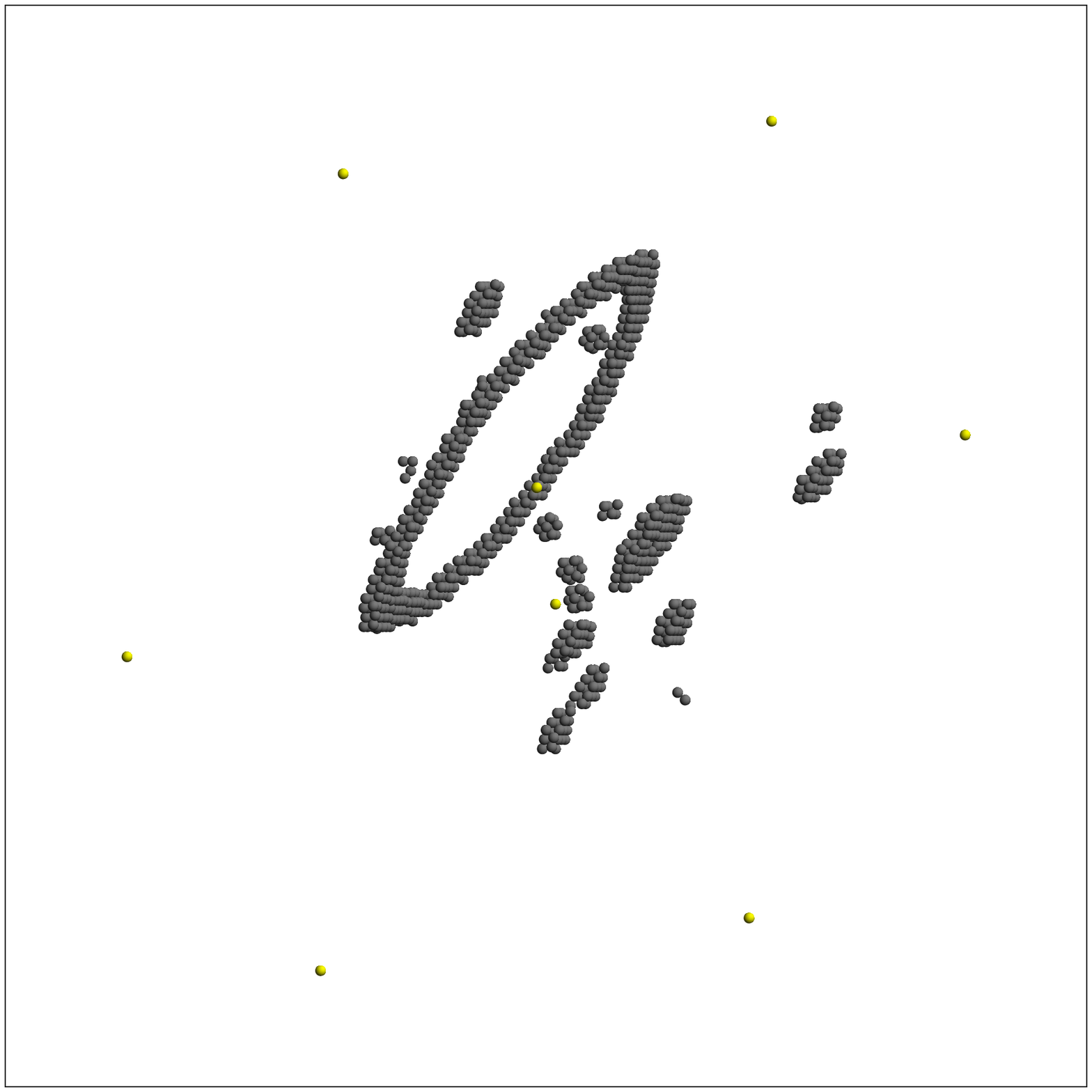, width=0.48\linewidth, clip=}\\[0.04\linewidth]
    \subfig{(c)}%
    \epsfig{file=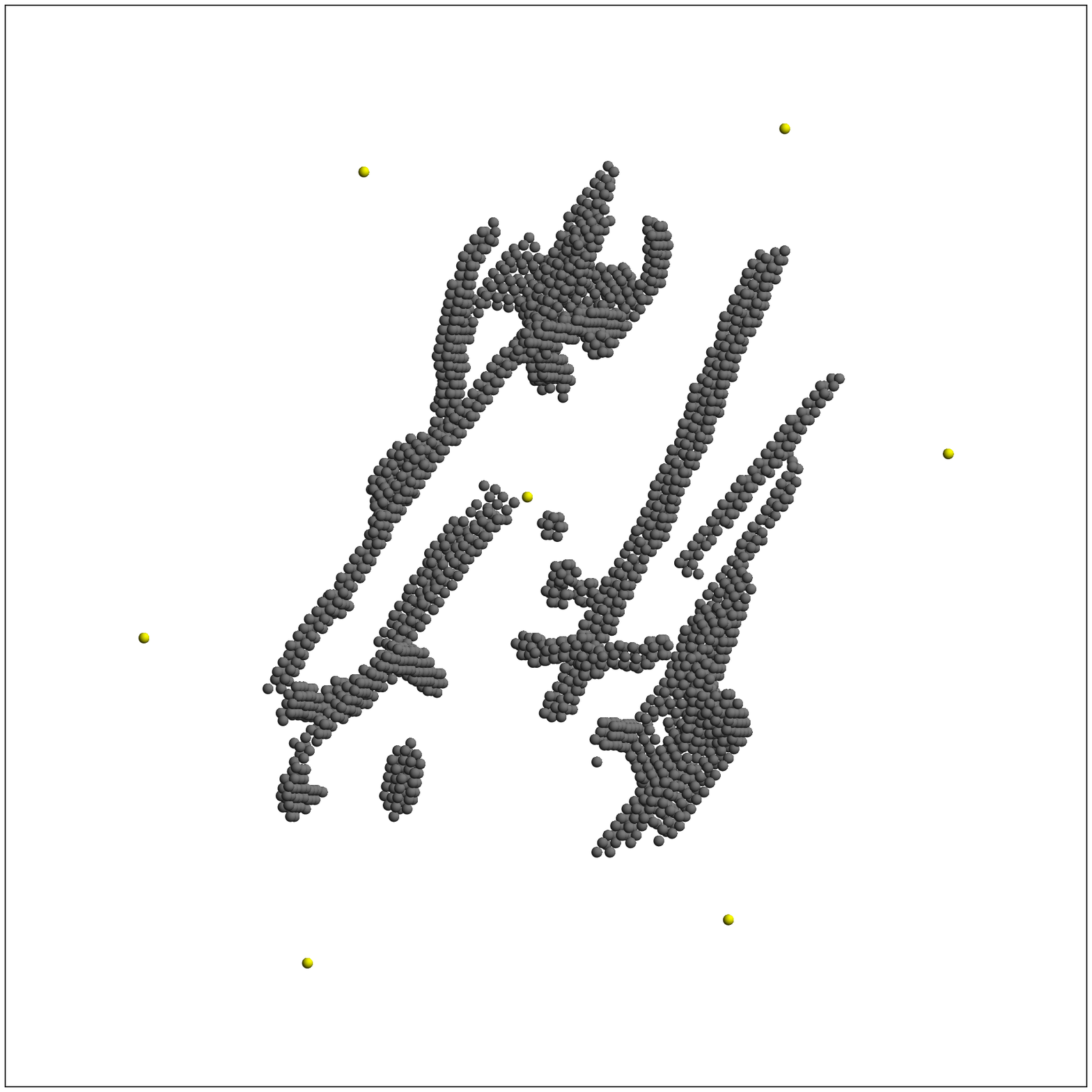, width=0.48\linewidth, clip=}\hfill
    \subfig{(d)}%
    \epsfig{file=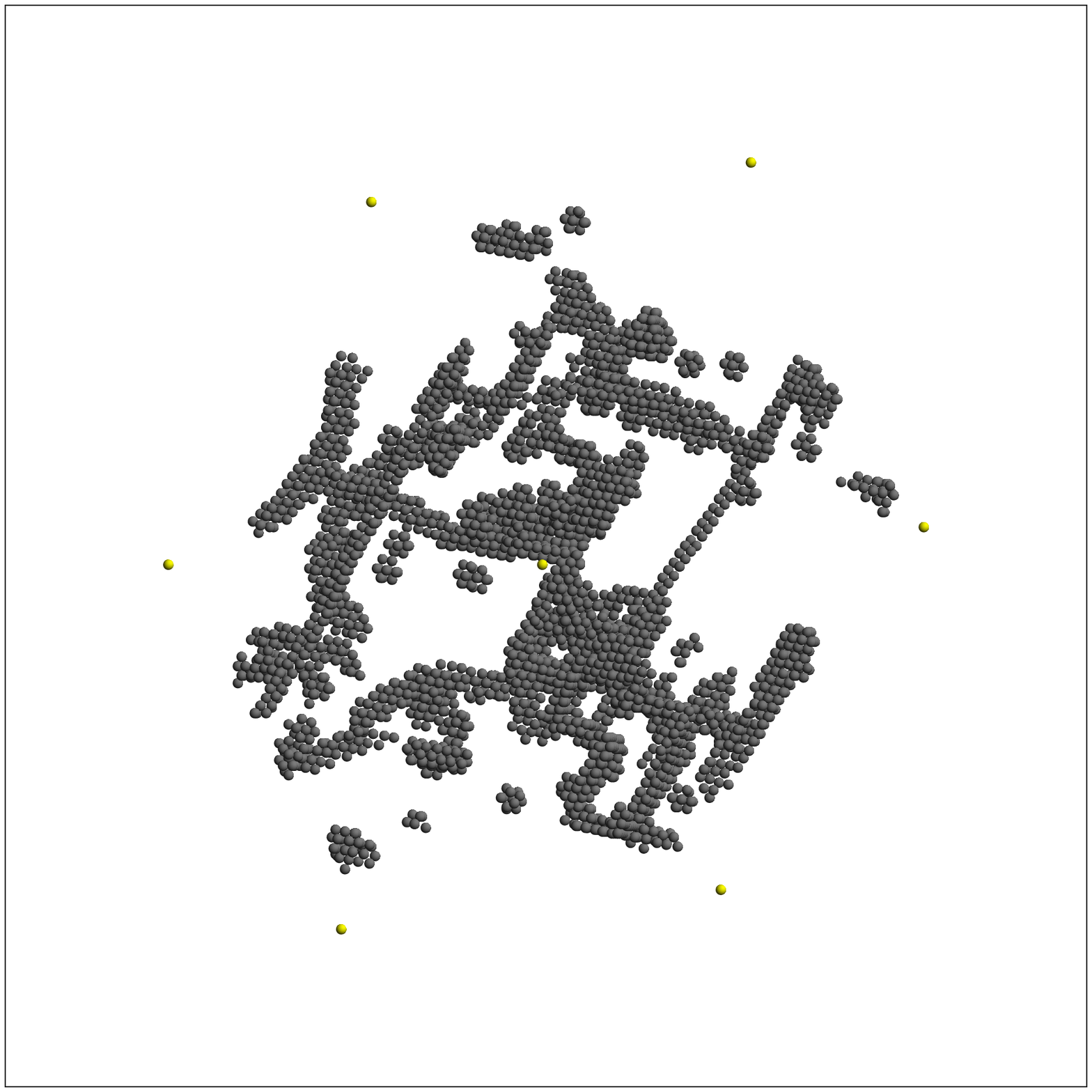, width=0.48\linewidth, clip=}
    \caption{System I.  In this and the following
      figures only atoms near crystal defects are shown.  Eight
      lighter gray atoms indicate the corners of the simulation cell.
      In \emph{part (a)} the four initial vacancies are seen.  The
      vacancies themselves are not shown, but the 12 atoms next to the
      vacancies are shown, as they are missing a nearest neighbor and
      therefore have the coordination number 11.  In \emph{part (b)}
      the first dislocation activity is seen.  \emph{Part (c)} shows
      the first generation of vacancies, and \emph{part (d)} shows the
      final configuration, where a high density of vacancies is seen.}
    \label{fig:sysI}
  \end{center}
\end{figure}

Figure \ref{fig:sysI} shows the deformation of the smaller system.  In
figure \ref{fig:sysI}(b) the first dislocation activity appears in the
form of a few dislocation loops.  The loops are nucleated
homogeneously, i.e.\ without the presence of dislocation sources or
crystal defects.  The exact time and position of the nucleation event
is determined by random thermal fluctuations.  The loops are faulted
loops consisting of a single partial dislocation with a stacking fault
inside the loop.  The loops expand rapidly.  Due to the periodic
boundary conditions the dislocations cannot disappear from the sample.
They will continue to move causing significant plastic deformation.
As the dislocations on different glide planes interact, dislocation
reactions occur and a number of triple junctions appear.  These
typically consist of two Shockley partials that meet to form a
stair-rod dislocation.  Continuous formation and destruction of
these sessile dislocations are seen in the simulations.

The dislocations observed in the simulations are mainly single
Shockley partials.  The homogeneous dislocation nucleation mechanism
makes it very difficult to nucleate the second partial.  Dislocations
are only nucleated at extremely high stresses, but as soon as the
first partial dislocation is created, it locally shields the stress
field and prevents the second partial from nucleating.  It probably
also plays a role that the stacking fault energy is
low.\footnote{Copper has a relatively low stacking fault energy, and
  furthermore the stacking fault energy in the simulation is
  approximately a factor of two below the experimental value (30
  mJ/m$^2$ versus 52 mJ/m$^2$).}  At high stresses the energy of the
stacking faults become small compared to other energies in the system,
such as the energy released by a moving dislocation.

After some dislocation activity has occurred we observe the first
creation of vacancies in figure \ref{fig:sysI}(c).  A ``sausage-like''
string of vacancies has appeared in the middle of the system.
Further dislocation activity breaks up the string of vacancies into
smaller clusters.

\begin{figure}[tp]
  \begin{center}
    \epsfig{file=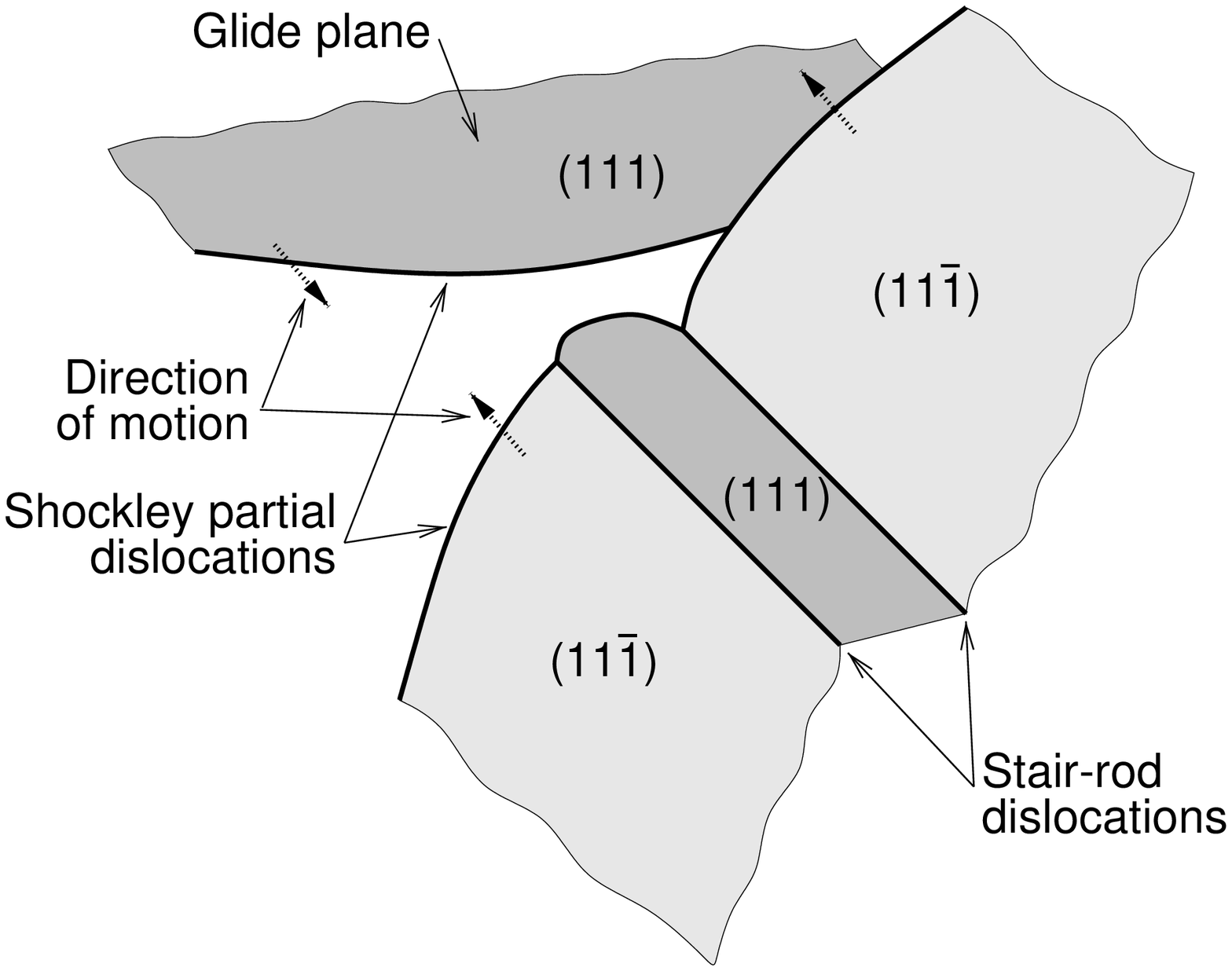, width=0.48\linewidth, clip=}\hfill
    \epsfig{file=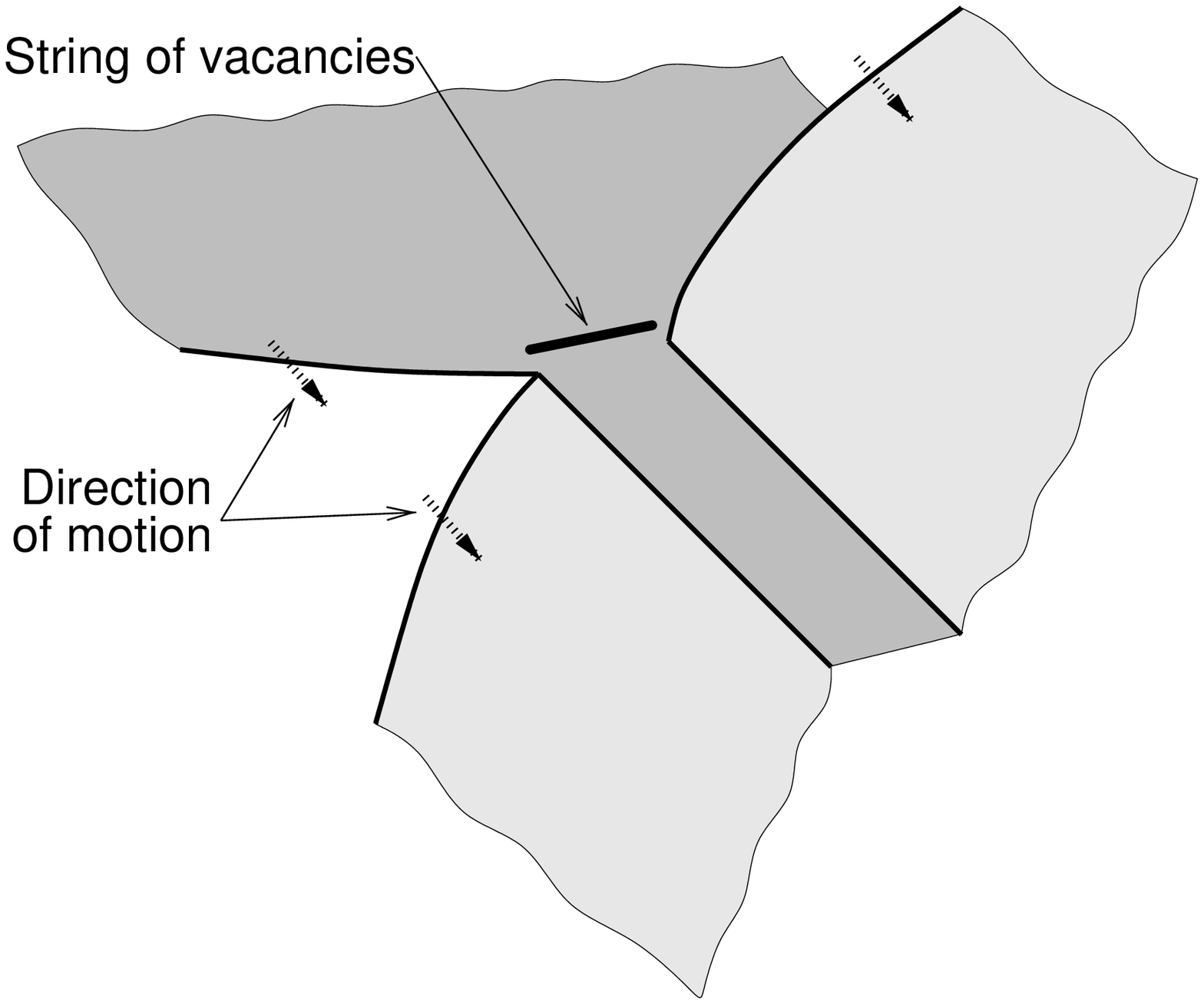, width=0.48\linewidth, clip=}
    \caption{The vacancies are generated by the annihilation of two
      segments of edge dislocations.  \emph{Left:} The
      configuration just before the annihilation.  A short segment of
      an edge dislocation is meeting an edge dislocation with the
      opposite Burgers vector.  They are gliding on adjacent (111)
      planes.  One of the dislocations is switching glide plane at two 
      point along the dislocation line, as the dislocation is moving
      stair-rod dislocations are left behind.  \emph{Right:} The
      dislocations have collided and a small dislocation loop has been 
      separated from the rest.  The dislocation loop quickly collapses 
      to a string of vacancies.}
    \label{fig:process}
  \end{center}
\end{figure}
To analyse the process by which such a cluster of vacancies is
generated, we repeated the relevant part of the simulation, using a
higher time resolution for the analysis of the process.  This revealed
the detailed process generating these vacancies.  The mechanism is
illustrated in figure \ref{fig:process}.  Dislocation interactions
have created dislocations where segments are on different glide
planes.  This makes it possible for a segment of a dislocation to be
annihilated.  If two edge dislocations gliding on adjacent glide
planes annihilate, a string of vacancies is formed.\footnote{If the
  dislocations had the opposite signs, a string of interstitials
  should be formed.  This is not observed in the simulations, probably
  because the energy of such a string is too high, so the dislocations
  pass each other instead of annihilating.}  This appears to be the
main source of vacancies in the simulations.

At the end of the simulation (figure \ref{fig:sysI}(d)) a large number
of vacancies have been formed, apparently by this mechanism.

\begin{figure}[tbp]
  \begin{center}
    \epsfig{file=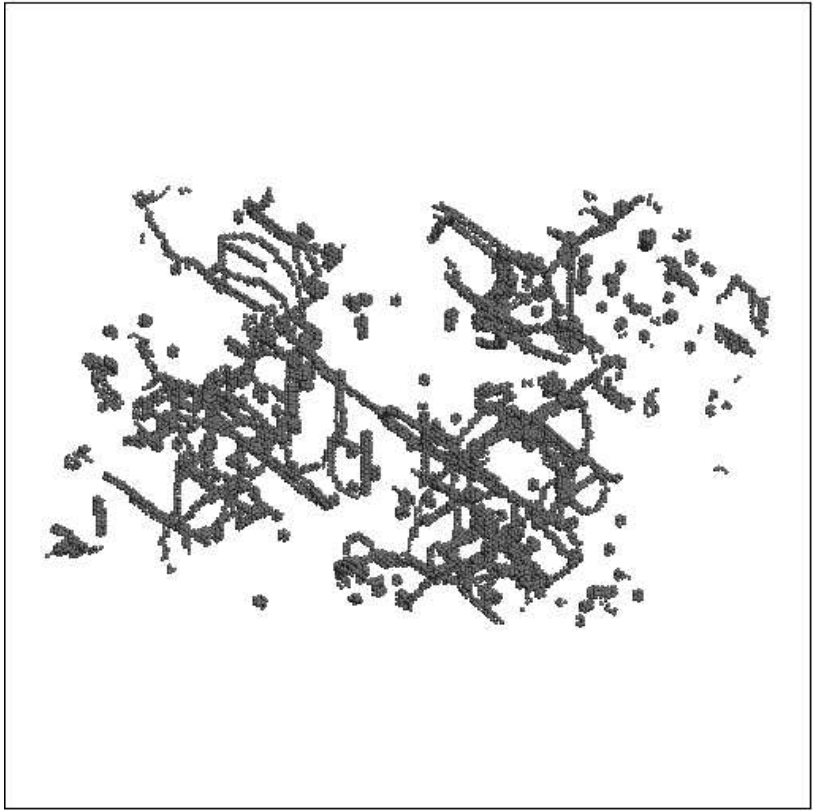, angle=-90, width=0.75\linewidth}
    \caption{The larger system after 40\% deformation.  Only a slice
      half the system size wide is shown.  Both dislocations and
      vacancies are clearly seen.}
    \label{fig:sysII}
  \end{center}
\end{figure}
The larger system behaves in a similar fashion.  Figure
\ref{fig:sysII} shows the larger system after the deformation.  As a
large number of dislocations were present, it  would have
been very difficult to see anything in the figure if the entire
system were plotted.  For this reason only a slice
through the system is shown.  Again a large number of both dislocations 
and vacancies are seen.

\section{Discussion}
\label{sec:discussion}

The simulations presented here indicate that the deformation of fcc
metals at ultra-high strain rate occurs by mechanisms that are not
very different from what is seen at more moderate strain rates.  Even
in a situation where there are no preexisting dislocations, and no
dislocation sources, dislocations are nucleated and are the main
carriers of the deformation.  The main difference between the
deformation mechanism in the simulations and in slowly deforming
materials is how the dislocations are nucleated.  This difference may
not be present in a real physical sample under ultra-high-speed
deformation, since dislocation sources will be present, even though it
is possible that other types of dislocation sources are activated
under high strain rate conditions.

The high concentration of vacancies in the simulations is due to the
very high dislocation density during the high-speed deformation.  As
the vacancies are generated when opposite-signed dislocations collide,
it must be expected that the generation rate depends on the dislocation
densities obtained and not just on the total deformation.  It is
therefore expected that much higher vacancy densities should be
observed experimentally after ultra-rapid deformation than
after deformation with moderate strain rate.

In the simulations presented here there are no free surfaces, and no
efficient way of removing dislocations except by annihilation.  We
therefore expect a higher dislocation density and a higher vacancy
production rate than those which would be observed with a more
realistic geometry.

To test the effect of free surfaces, we have made a similar simulation
with four free surfaces (i.e.\ the system has the form of a rod being
pulled).  In this case dislocations were nucleated at the corners,
passed through the system and disappeared at the free surfaces on the
other side of the sample.  Only a few dislocations were moving through
the system at the same time.  No vacancies were generated, as the
dislocations never collided.  Preliminary simulations with two free
surfaces (i.e. foil geometry) indicate that this geometry behaves much
like the bulk simulations, as dislocations perpendicular to the
surfaces cannot move to the surface and disappear.  A large
dislocation density builds up, and vacancies are generated.

\section{Conclusions}
\label{sec:concl}

Atomic-scale computer simulations using molecular dynamics indicate
that ultra-high-speed deformation of fcc copper occurs by the
nucleation and motion of dislocations.  Even in the absence of
dislocation sources, dislocations will be nucleated at sufficiently
high stresses.

During the deformation, a large number of vacancies are formed, in
agreement with experimental investigations of ultra-high strain rate
deformation of thin foils of fcc metals \cite{KiSaKiArOgArSh99}.  The
vacancies are produced when segments of edge dislocations moving on
adjacent slip planes annihilate.  This generates ``strings'' of
dislocations.  Further dislocation activity tends to break these
strings into smaller clusters.  Finally, over time spans much longer
than those which we can simulate using molecular dynamics, the
vacancies will diffuse and will eventually cluster, forming stacking
fault tetrahedra.

\section{Acknowledgments}

We would like to thank Prof.\ Kiritani for fruitful discussions.
Center for Atomic-scale Materials Physics (CAMP) is sponsored by the
Danish National Research Foundation.  This work was done as a
collaboration between CAMP and the Engineering Science Center for
Structural Characterization and Modelling of Materials at Materials
Research Department, Ris{\o}.  Parallel computer time was financed by
the Danish Research Councils through grant no.~9501775.


\begin{thebibliography}{1}
\expandafter\ifx\csname url\endcsname\relax
  \def\url#1{\texttt{#1}}\fi
\expandafter\ifx\csname urlprefix\endcsname\relax\def\urlprefix{URL }\fi

\bibitem{KiSaKiArOgArSh99}
M.~Kiritani, Y.~Satoh, Y.~Kizuka, K.~Arakawa, Y.~Ogasawara, S.~Arai, and
  Y.~Shimomura, \emph{Phil. Mag. A} \textbf{79}, 797 (1999).

\bibitem{ScDiJa98}
J.~Schi{\o}tz, F.~D. Di~Tolla, and K.~W. Jacobsen, \emph{Nature} \textbf{391},
  561 (1998).

\bibitem{ScVeDiJa99}
J.~Schi{\o}tz, T.~Vegge, F.~D. Di~Tolla, and K.~W. Jacobsen, \emph{Phys. Rev.
  B} \textbf{60}, 11971 (1999).

\bibitem{SwCa98}
H.~Van~Swygenhoven and A.~Caro, \emph{Phys. Rev. B} \textbf{58}, 11246 (1998).

\bibitem{JaNoPu87}
K.~W. Jacobsen, J.~K. N{\o}rskov, and M.~J. Puska, \emph{Phys. Rev. B}
  \textbf{35}, 7423 (1987).

\bibitem{JaStNo96}
K.~W. Jacobsen, P.~Stoltze, and J.~K. N{\o}rskov, \emph{Surf. Sci.}
  \textbf{366}, 394 (1996).

\bibitem{AlTi87}
M.~P. Allen and D.~J. Tildesley, \emph{Computer simulation of liquids}
  (Claredon Press, Oxford, 1987).

\bibitem{JoAn88}
H.~J\'onsson and H.~C. Andersen, \emph{Phys. Rev. Lett.} \textbf{60}, 2295
  (1988).

\bibitem{ClJo93}
A.~S. Clarke and H.~J\'onsson, \emph{Phys. Rev. E} \textbf{47}, 3975 (1993).

\end{thebibliography}

\end{document}